\def\year{2018}\relax
\documentclass[letterpaper]{article} 
\usepackage{aaai18}  
\usepackage{times}  
\usepackage{helvet}  
\usepackage{courier}  
\usepackage{url}  
\usepackage{graphicx}  
\usepackage{booktabs}
\usepackage{graphics}
\usepackage{blindtext}
\usepackage{siunitx}
\usepackage{multirow}
\usepackage[justification=centering]{caption}
\begin{document}
%
\title{You Are Your Metadata: Identification and Obfuscation of Social Media Users using Metadata Information}
\author{Beatrice Perez$^{1}$, Mirco Musolesi$^{1,2}$, Gianluca Stringhini$^{1}$\\
  $^{1}$University College London, London, UK\\
  $^{2}$The Alan Turing Institute, London, UK\\
  \{beatrice.perez.14, m.musolesi, g.stringhini\}@ucl.ac.uk
}
\maketitle
\begin{abstract}
Metadata are associated with most of the information we produce in our daily interactions and communication in the digital world. Yet, surprisingly, metadata are often still categorized as non-sensitive. Indeed, in the past, researchers and practitioners have mainly focused on the problem of the identification of a user from the \textit{content} of a message.

In this paper, we use Twitter as a case study to quantify the uniqueness of the association between metadata and user identity and to understand the effectiveness of potential obfuscation strategies. More specifically, we analyze atomic fields in the metadata and systematically combine them in an effort to classify new tweets as belonging to an account using different machine learning algorithms of increasing complexity.  We demonstrate that, through the application of a supervised learning algorithm, we are able to identify any user in a group of 10,000 with approximately 96.7\% accuracy.  Moreover, if we broaden the scope of our search and consider the  10 most likely candidates we increase the accuracy of the model to 99.22\%. We also found that data obfuscation is hard and ineffective for this type of data: even after perturbing 60\% of the training data, it is still possible to classify users with an accuracy higher than 95\%. These results have strong implications in terms of the design of metadata obfuscation strategies, for example for data set release, not only for Twitter, but, more generally, for most social media platforms. 
\end{abstract}

\section{Introduction}

Platforms like Facebook, Flickr, and Reddit allow users to share links, documents, images, videos, and thoughts.
Data has become the newest form of currency and analyzing data is both a business and an academic endeavor. 
When online social networks (OSNs) were first introduced, privacy was not a major concern for users and therefore not a priority for service providers. With time, however, privacy concerns have risen: users started to consider the implications of the information they share~\cite{silentListeners,humphreys2010much} and in response OSN platforms have introduced coarse controls for users to manage their data~\cite{hummingbird}. 
Indeed, this concern is heightened by the fact that this descriptive information can be actively analyzed and mined for a variety of purposes, often beyond the original design goals of the platforms. For example, information collected for targeted advertisement might be used to understand political and religious inclinations of a user. The problem is also exacerbated by the fact that often these datasets might be publicly released either as part of a campaign or through information leaks.

Previous work shows that the content of a message posted on an OSN platform reveals a wealth of information about its author. Through text analysis, it is possible to derive age, gender, and political orientation of individuals~\cite{rao2010classifying}; the general mood of groups~\cite{groupEmotions} and the mood of individuals~\cite{IndividualEmotions}. Image analysis reveals, for example, the place a photo was taken~\cite{im2gps}, the place of residence of the photographer~\cite{hometown}, or even the relationship status of two individuals~\cite{shoshitaishvili2015portrait}. If we look at mobility data from location-based social networks, the check-in behavior of users can tell us their cultural background~\cite{eatANDdrink} or identify users uniquely in a crowd~\cite{check-in}. Finally, even if an attacker only had access to anonymized datasets, by looking at the structure of the network someone may be able to re-identify users ~\cite{deanonymizing}. Most if not all of these conclusions could be considered privacy-invasive by users, and therefore the content is what most service providers are starting to protect. However, access control lists are not sufficient. We argue that the behavioral information contained in the metadata is just as informative. 

Metadata has become a core component of the services offered by OSNs. For example, Twitter provides information on users mentioned in a post, the number of times a message was re-tweeted, when a document was uploaded, and the number of interactions of a user with the system, just to name a few. These are not merely extra information: users rely on these to measure the credibility of an account~\cite{wang2012social} and much of the previous research in fighting social spam relies on account metadata for detection~\cite{benevenuto2010detecting,stringhini2010detecting}. 

In this paper, we present an in-depth analysis of the identification risk posed by metadata to a user account. We treat identification as a classification problem and use supervised learning algorithms to build \textit{behavioral signatures} for each of the users. Our analysis is based on metadata associated to micro-blogging services like Twitter: each tweet contains the metadata of the post as well as that of the account from which it was posted. However, it is worth noting that the methods presented in this work are generic and can be applied to a variety of social media platforms with similar characteristics in terms of metadata. In that sense, Twitter should be considered only as a case study, but the methods proposed in this paper are of broader applicability.
The proposed techniques can be used in several practical scenarios such as when the identifier of an account changes over time, when a single user creates multiple accounts, or in the detection of legitimate accounts that have been hijacked by malicious users.

In security, there are at least two areas that look at identity from opposite perspectives: on one hand, research in authentication looks for methods that, while unobtrusive and usable, consistently identify users with low false positive rates~\cite{continuousAuth}; and, on the other hand, work on obfuscation and differential privacy aims to find ways by which we can preserve an individual's right to privacy by making information about them indistinguishable in a set~\cite{tclose}. This study is relevant to both: we claim that in the same way that our behavior in the physical world is used to identify us~\cite{Bailey201477,SilentSense,wang2009behavioral}, 
the interactions of a user with a system, as represented by the metadata generated during the account creation and its subsequent use, can be used for identification. 
If this is true, and metadata can in fact be linked to our identity and as it is seldom if ever protected, it constitutes a risk for users' privacy.  
Our goal is therefore, to determine if the information contained in users' metadata is sufficient to fingerprint an account.  
Our contributions can be summarized as follows:
\begin{itemize}
\item We develop and test strategies for user identification through the analysis of metadata through state-of-the-art machine learning algorithms, namely Multinomial Logistic Regression (MLR)~\cite{bishop}, Random Forest (RF)~\cite{randomForest}, and K-Nearest Neighbors (KNN)~\cite{classifier}. 
\item We provide a performance evaluation of different classifiers for multi-class identification problems, considering a variety of dimensions and in particular the characteristics of the training set used for the classification task.
\item We assess the effectiveness of two obfuscation techniques in terms of their ability of hiding the identity of an account from which a message was posted.
\end{itemize}

\section{Motivation}
\begin{table*}
\centering
\caption{Description of relevant data fields.}
\label{input_description}
\begin{tabular}{ll}
\toprule
\textbf{Feature} & \textbf{Description}\\ \midrule
Account creation & UTC time stamp of the account creation time.\\ 
Favourites count & The number of tweets that have been marked as `favorites' of this account.\\
Follower count & The number of users that are following this account.\\
Friend count & The number of users this account is following.\\
Geo enabled & (boolean) Indicates whether tweets from this account is geo-tagged.\\
Listed count  & The number of public lists that include the account.\\
Post time stamp & UTC time of day stamp at which the post was published.\\
Statuses count & The number of tweets posted by this account.\\
Verified & (boolean) Indicates that Twitter has checked the identity of the user that owns this account.\\
\bottomrule
\end{tabular}
\end{table*}

\subsection{Formal Definition of the Study}
We consider a set of users 
\begin{center}
$\mathbf{U} =\{ u_1, u_2, \dots, u_k, \dots , u_M\}$. 
\end{center}
Each user $u_i$ is characterized by a finite set of features 
\begin{center}
$\mathbf{X^{u_k}}=\{ {x^{u_k}}_1, {x^{u_k}}_2, \dots , {x^{u_k}}_R\} $. 
\end{center}
In other words, we consider $M$ users and each user is represented by means of $R$ features. Our goal is to map this set of users to a set of identities 
\begin{center}
$\mathbf{I}=\{ i_1, i_2, \dots , i_k, \dots, i_M\}$. 
\end{center}
We assume that each user $u_k$ maps to a unique identity $i_l$.

Identification is framed in terms of a classification problem: in the training phase, we build a model with a known dataset; in our case, we consider a dataset in which a user $u_k$ as characterized by features $X_{u_k}$, extracted from the user's profile, is assigned an identity $i_l$. Then, in the classification phase, we assign to each user $\hat{u}_k$, for which we assume that the identity is unknown, a set of probabilities over all the possible identities.

More formally, for each user $\hat{u}_k$ (i.e., our test observation) the output of the model is a vector of the form 
\begin{center}
$\mathbf{P}_{\mathbf{I}}(\hat{u}_k)=\{p_{i_1}(\hat{u}_k), p_{i_2}(\hat{u}_k), \dots , p_{i_M}(\hat{u}_k)\}$, 
\end{center}

\noindent where $p_{i_1}(\hat{u}_k)$ is the probability that the test observation related to $\hat{u}_k$ will be assigned to identity $i_1$ and so on. We assume a \textit{closed} system where all the observations will be assigned to users in $\mathbf{I}$, and therefore, $\sum_{i_{l} \in \mathbf{I}} p_{i_l}(\hat{u}_k) = 1 $. Finally, the identity assigned to the observation will be the one that corresponds to $ argmax_{i_l} \{ p_{i_l}(\hat{u}_k)\}$.

Two users with features having the same values are indistinguishable. 
Moreover, we would like to point out that the values of each feature can be static (constant) over time or dynamic (variable) over time. An example of a static feature is the account creation time. An example of a dynamic feature is the number of followers of the user at the time the tweet was posted. 
Please also note that, from a practical point of view, our objective is to ascertain the identity of a user in the test set. In the case of a malicious user, whose identity has been modified over time, we assume that the `real' identity is the one that is found in the training set. In this way, our method could also be used to group users with very similar characteristics and perhaps conclude that they belong to the same identity.

\subsection{Attack Model}
The goal of the study is to understand if it is possible to correctly identify an account given a series of features extracted from the available metadata. In our evaluation, 
as discussed before,
the input of the classifier is a set of new (unseen) tweets. We refer to a successful prediction of the account identity as a \textit{hit} and an unsuccessful one as a \textit{miss}. We assume that the attacker is able to access the metadata of tweets from a group of users together with their identities (i.e., the training set) and that the new tweets belong to one of the users in the training set.

We present the likelihood of success of an identification attack where the adversary's ultimate goal is to identify a user from a set given this knowledge about the set of accounts. To achieve this, we answer this question: \textit{Is it possible to identify an individual from a set of metadata fields from a randomly selected set of Twitter user accounts?}
 
\section{Methods}

\subsection{Metadata and the case of Twitter}
We define metadata as the information available pertaining to a Twitter post. This is information that describes the context in which the post was shared. Apart from the 140 character message, each tweet contains about 144 fields of metadata. Each of these fields provides additional information about: the account from which it was posted; the post (e.g., time, number of views); other tweets contained within the message; various entities (e.g., hashtags, URLs, etc); and the information of any users directly mentioned in it. From these features (in this work we will use features, fields, inputs to refer to each of the characteristics available from the metadata) we created combinations from a selection of 14 fields as a basis for the classifiers. 

\subsection{Feature Selection}
Feature selection methods can be essentially grouped in three classes following the classification proposed by Liu and Yu in~\cite{featureSelection}: the filter approach that ranks features based on some statistical characteristics of the population; the wrapper approach that creates a rank based on metrics derived from the measurement algorithm; and a group of hybrid methods which combine the previous two approaches. In the same paper, the authors claim that the wrapper method is guaranteed to find the optimal combination of inputs for classification based on the selected criteria. Three years later, in~\cite{classifier}, Huang et al. provided validation by experimentally showing that for classification, the wrapper approach results in the best possible performance in terms of accuracy for each algorithm. 

From the three proposed, the only method that allows for fair comparison between different algorithms is the wrapper method. Since it guarantees optimal feature combination on a per algorithm basis, it eliminates any bias in the analysis due to poor selection. Ultimately, we conducted a comprehensive stratified search over the feature space and obtained a ranking per level for each of the algorithms. Here, a level corresponds to the number of features used as input for the classifier and we will use $n$ to denote it. In the first level, where $n=1$ we looked at the predictive power of each of the 14 features individually; for $n=2$ we looked at all combinations of pairs of features, and so on. We use the term combinations to describe any group of $n$ un-ordered features throughout the paper.

The features selected were those that describe the user account and were not under direct control of the user with the exception of the account ID which was excluded as it was used as ground truth (i.e., label) of each observation. As an example, the field describing the users' profile background color was not included in the feature list while the number of friends and number of posts were. Table \ref{input_description} contains a description of the fields selected.

\subsection{Implementation of the Classifiers}
\label{ImplementationOftheClassifier}
We consider three state-of-the-art classification methods: Multinomial Logistic Regression (MLR)~\cite{bishop}, Random Forest (RF)~\cite{randomForest}, and K-Nearest Neighbors (KNN)~\cite{classifier}. 
Each of these algorithms follows a different method to make the recommendation and they are all popular within the community. 

We use the implementation of the algorithms provided by sci-kit learn~\cite{scikit}, a Python library. The optimization of the internal parameters for each classifier was conducted as a combination of best practices in the field and experimental results in which we used the cross-validated grid search capability offered by SciKit-learn. In summary, we calculated the value of the parameters for each classifier as follows. For KNN, we consider the single closest value based on the Euclidean distance between the observations; for RF, we chose entropy as the function to measure the effectiveness of the split of the data in a node; finally, for MLR, we selected the limited-memory implementation of the Broyden-Fletcher-Goldfarb-Shanno (\textit{LM-BFGS}) optimizer as the value to optimize~\cite{Liu1989}.

\subsection{Obfuscation and Re-Identification}
\label{methods-obfuscation}
Obfuscation can only be understood in the context of data sharing. The goal of obfuscation is to protect the private individual fields of a dataset by providing only the result of some function computed over these fields~\cite{cloud2012}. 
To succeed, the possibilities are either to obfuscate the data or develop algorithms that protect it~\cite{pdmMalik}. We reasoned that an attacker will have access to any number of publicly available algorithms, this is outside of our control. However, we could manipulate the granularity of the information made available. Our task is to determine whether doing so is an effective way of protecting user privacy, particularly when obfuscated metadata is released. 

In this work, we focus on two classic obfuscation methods: data randomization and data anonymization~\cite{ppdmAgrawal,desensitization,collaborative}. Data anonymization is the process by which the values of a column are grouped into categories and each reading is replaced by an index of its corresponding category. Data randomization, on the other hand, is a technique that alters the values of a subset of the data points in each column according to some pre-determined function.  We use rounding as the function to be applied to the data points. For each of the values that were altered, we rounded to one less than the most significant value (i.e., 1,592 would be 1,600 while 31 would be 30). We measured the level of protection awarded by randomization by recording the accuracy of the predictions as we increased the number of obfuscated data points in increments of 10\% until we reached full anonymization (i.e., 100\% randomization) of the training set.

\subsection{Inference Methods}
Statistical inference is the process by which we generalize from a sample a characteristic of the population. Bootstrapping is a computational method that allows us to make inferences without making any assumptions about the distribution of the data and without the need of formulas to describe the sampling process. With bootstrapping we assume that each sample is the population and then aggregate the result from a large number of runs (anywhere between 50 and 1,000 times depending on the statistic being drawn)~\cite{bootstrapping}. In this study, we are primarily interested in the precision and accuracy of each classifier as a measure of their ability to predict the correct user given a tweet.  
The results we present in the paper are an average over 200 repetitions of each experiment. In each experiment, the input data was randomly split between training and testing sets using a 7:3 proportion, which is a typical setting in the evaluation of machine learning algorithms.

\section{Experimental Settings}
\begin{table}
\caption{KNN classification accuracy using ten observations per user using follower count and friend count as features in input. We ran each experiment for an increasing number of users $u$.}
\label{user_no}
\centering
\resizebox{.7\columnwidth}{!}{%
\begin{tabular}{ccc} \toprule
$ u $ & top result & top 5 \\ \midrule
10 & 94.283 ($\pm 0.696$) & 98.933 ($\pm 0.255$) \\ 
100 & 86.146 ($\pm 0.316$) & 96.770 ($\pm 0.143$) \\ 
1,000 & 70.348 ($\pm 0.112$) & 90.867 ($\pm 0.076$)\\ 
10,000 & 47.639 ($\pm 0.039$) & 76.071 ($\pm 0.029$)\\ 
100,000 & 28.091 ($\pm 0.089$) & 55.438 ($\pm 0.192$) \\ 
\hline
\bottomrule
\end{tabular}%
}
\end{table}
\subsection{Dataset}
For data collection, we used the Twitter Streaming Public API~\cite{SearchAPI}. Our population is a random\footnote{It is worth noting that since we use the public Twitter API we do not have control on the sampling process. Having said that, an attacker will most probably access the same type of information. A typical use case is the release of data set of tweets usually obtained in the same way.} sample of the tweets posted between October 2015 and January 2016 (inclusive).
During this period we collected approximately 151,215,987 tweets corresponding 11,668,319 users. However, for the results presented here, we considered only users for which we collected more than 200 tweets. Our final dataset contains tweets generated by 5,412,693 users. 

\subsection{Ethics Considerations}
Twitter is the perfect platform for this work. On one hand, users posting on Twitter have a very low expectation of privacy: it is in the nature of the platform to be open and to reach the widest audience possible. Tweets must always be associated with a valid account and, since the company does not collect demographic information about users upon registration, the accounts are not inherently linked to the physical identity of the users. Both these factors reduce but not eliminate any ethical concerns that may arise from this work. 
Nonetheless, we submitted this project for IRB approval and proceeded with their support.

.\subsection{Experimental Variables}
\label{expVar}

\subsubsection{Number of Users}
\label{noUser}

\begin{figure}
\caption{Change in accuracy for a single feature combination and increasing number of users.}
\label{sample_u}
\centering
\resizebox{.8\columnwidth}{!}{%
\includegraphics[width=\linewidth]{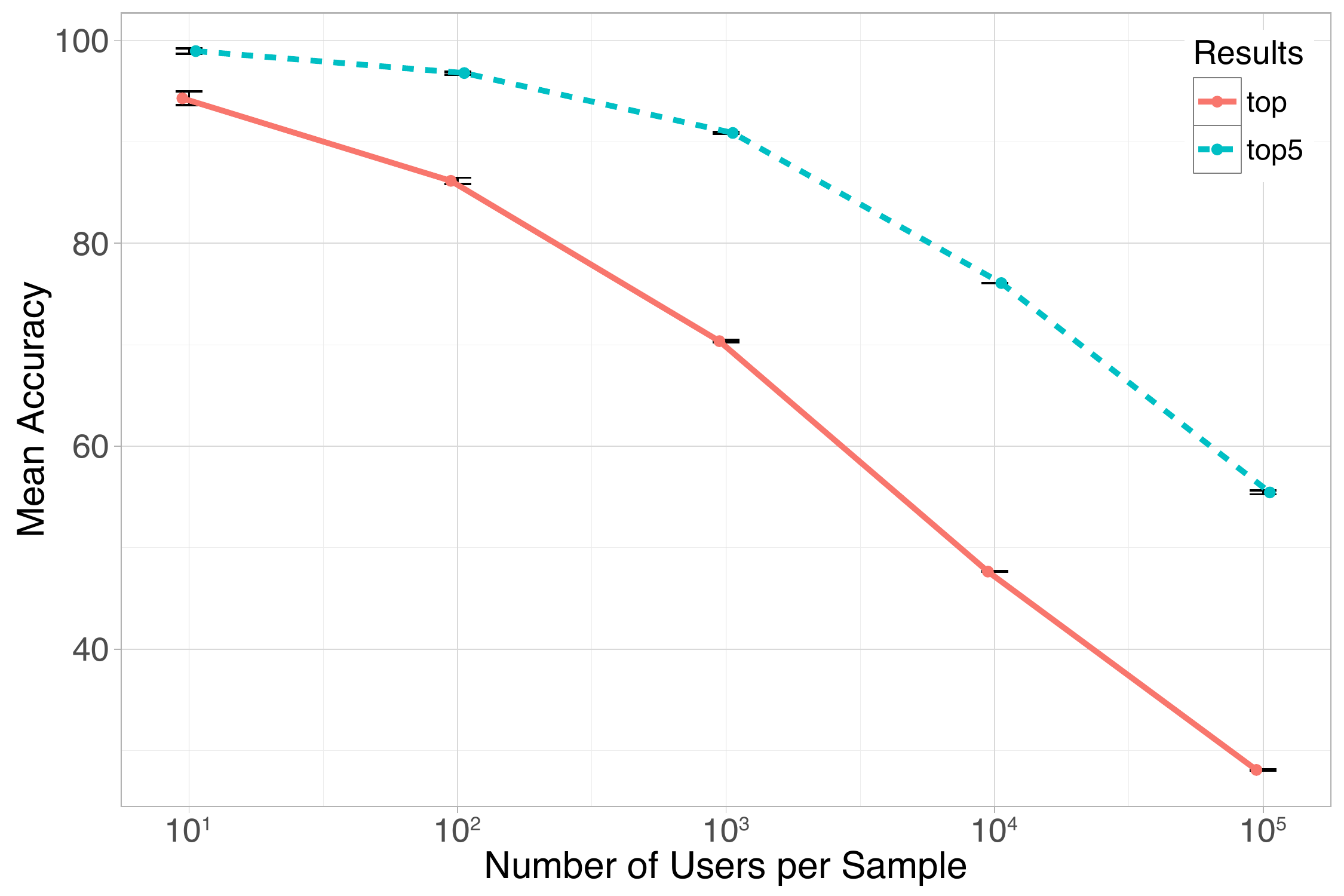}%
}
\end{figure}

As an attack, guessing is only viable for smaller user pools: indeed, there is a 1:10 probability of randomly classifying a tweet correctly for a user pool made up of 10 users, whereas there is a 1:10,000 probability in a pool of 10,000 users. The likelihood of guessing correctly is inversely proportional to the number of users. Therefore, the first task was to compare and describe the way in which increasing the number of users affects the accuracy of the classifiers.
We evaluate each algorithm (i.e., MLR, RF, KNN) on a specific configuration of training data in order to obtain a trained model.
We trained models for all feature combinations, however, we present only the results for the best combination of parameters for each classifier. 

We first analyze the impact of the number of classes.
In Figure \ref{sample_u} we present with fixed parameters the effect of increasing only the number of outputs for each of the classifiers. Each model was built with two input features (i.e., $n=2$ where the features are number of friends and number of followers) and 10 observations per user. 
Some of the results we present are independent of the underlying classification algorithm. For these instances, we present results using only KNN. As it will be shown later in the paper, KNN shows the best performance in terms of prediction and resource consumption.

Figure \ref{sample_u} shows that the loss in accuracy is at worst linear. In a group of 100,000 users, the number of friends and of followers was sufficient to identify 30\% of the users (around 30,000 times better than random). However, while the accuracy of the classification gradually declines, there is a dramatic increase in the cost (in terms of time) when building the model.
As we will discuss in the last part of the Results section, the greatest obstacle with multi-class classification is the number of classes included in the model. 

\subsubsection{Number of Entries per User} 
\label{entriesPerUser}
The next variable we consider is the number of observations per user per model. Our objective is to visualize the relationship between accuracy and the number of observed tweets to set a minimum value for the rest of the study. 

\begin{figure}
\centering
\caption{Performance of the top 20\% of combinations per classifier for increasing observations per user.}
\label{entry_no}
\resizebox{.8\columnwidth}{!}{%
\includegraphics[width=\linewidth]{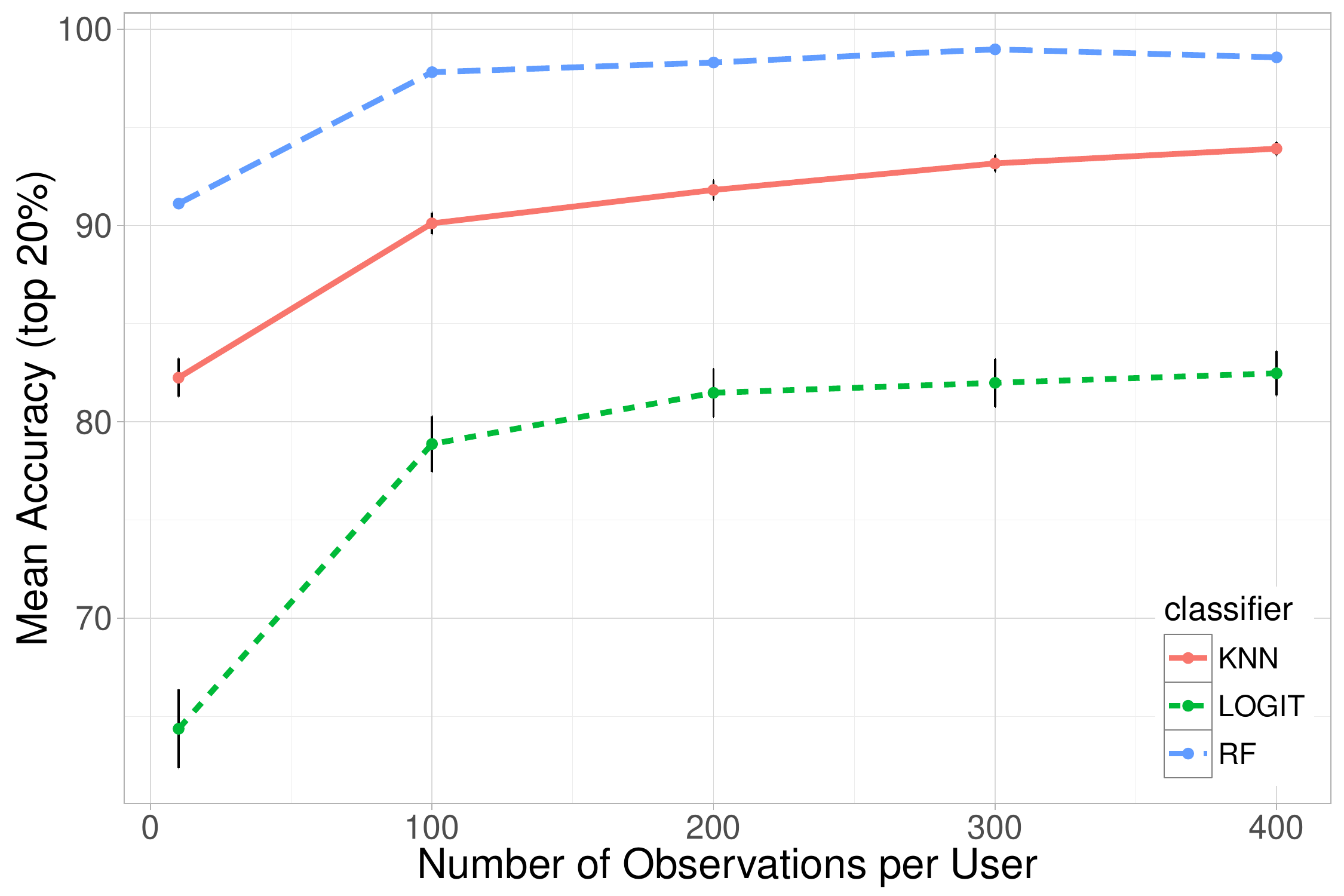}%
}
\end{figure}

To set this parameter, we fixed the number of input features at $n=2$ and $u=1,000$ then we ran models with 10, 100, 200, 300, and 400 tweets per user. Figure \ref{entry_no} shows the aggregated results for the 20\% most accurate feature combinations over 400 iterations. As the figure shows, 10 entries are not enough to build robust models. For all classifiers we see that the behavior is almost asymptotic. There is a significant increase in accuracy as the number of observations per user reaches 100. However, for all subsequent observations, the variation is less apparent. Each of the points in the graph contains the confidence interval associated with the measurement. However, for RF the largest error is 0.2.
It is worth noting that with our data set we could only scale as far as 10,000 users.

Each experiment presented in the remainder of the paper was repeated 200 times (each time with a different configuration in terms of users and number of tweets).
By standardizing the number of observations per user at 200 tweets, we preempt two problems: first, by forcing all users to have the same number of tweets we reduce the likelihood of any user not having a sufficient number of entries in the training set; and second, it prevents our results from being biased towards any user with a disproportionate number of observations. This might be considered as potentially artificial, but we believe it represents a realistic baseline for evaluating this attack.

\section{Results}
%
\subsection{Identification}
In building and comparing models we are interested in the effects and interactions of the number of variables used for building the models. We denote the number of input features with $n$ and the number of output classes which we refer to as $u$.

We define accuracy as the correct assignment of an observation to a user. In a multi-class algorithm, the predicted result for each observation is a vector containing the likelihood of that observation belonging to each of the users present in the model. A prediction is considered correct if the user assigned by the algorithm corresponds to the true account ID.
In the rest of the paper, we report aggregated results with 95\% confidence intervals. 

In our analysis, the account creation time was found to be highly discriminative between users. We present results with the dynamic features, defined below. Then we present our findings with the account creation time. Finally, we give a comprehensive analysis of the task of identification given combinations of both static and dynamic features for three different classifiers.

\subsubsection{Static Attribute: Account Creation Time}
\label{results-ACT}

\begin{table}[t]
\caption{Entropy calculation for feature list.}
\label{entropy_table}

\centering
\resizebox{.4\columnwidth}{!}{%
\begin{tabular}{cc} \toprule
feature & entropy \\ \midrule
ACT & 20.925  \\ 
statuses count & 16.132  \\ 
follower count & 13.097   \\ 
favorites count & 12.994  \\ 
friend count & 11.725  \\ 
listed count & 6.619  \\ 
second & 5.907 \\ 
minute & 5.907  \\ 
day & 4.949  \\ 
hour & 4.543 \\ 
month & 3.581  \\ 
year & 2.947 \\ 
post time & 1.789 \\ 
geo enabled & 0.995 \\ 
verified & 0.211 \\ 
\hline
\bottomrule
\end{tabular}%
}
\end{table}


Twitter includes the Account Creation Time (ACT) in every published tweet. The time stamp represents the moment in which the account was registered, and as such, this value is constant for all tweets from the same account. Each time stamp is composed of six fields: day, month, year, hour, minute, and second of the account creation.  
For perfect classification (i.e., uniqueness), we look for a variable whose value is constant within a class but distinct across classes 
(as explained before, each user is a class).
An example is the account creation time.
We tested the full ACT using KNN and found that, even for 10,000 users, classifying on this feature resulted in 99.98\% accuracy considering 200 runs. 
Nonetheless, the ACT is particularly interesting because while it represents one unique moment in time (and thus infinite and highly entropic), it is composed of periodic fields with a limited range of possible values. As we see in Table \ref{entropy_table}, each of the fields has at least a 75\% decrease in entropy as compared to the combined ACT. 
Since full knowledge of the ACT can be considered as the trivial case for our classification problem, in the following sections we will consider the contribution of each field of the ACT separately.

\subsubsection{Dynamic Attributes}
\begin{table}[t]
\caption{KNN Classification using dynamic features.}
\label{NAC-KNN}

\centering
\resizebox{.9\columnwidth}{!}{%
\begin{tabular}{cclc} 
\toprule
{$u$} & {$n$} & {features} & {accuracy} \\ \midrule

\multirow{4}{*}{10,000} & \multirow{2}{*}{3} & friend, follower, listed count & 92.499 ($\pm 0.0008$) \\
&  & friend, follower, favorite count & 91.158 ($\pm 0.0006$) \\
& \multirow{2}{*}{2} & friend, follower count & 83.721 ($\pm 0.0005$) \\
&  & friend, favorite count & 78.547 ($\pm 0.0006$) \\ \midrule

\multirow{4}{*}{1,000} & \multirow{2}{*}{3} & friend, follower, listed count & 95.702 ($\pm 0.0037$) \\
&  & friend, follower, favorite count & 93.474 ($\pm 0.0015$) \\
& \multirow{2}{*}{2} & friend, follower count & 91.565 ($\pm 0.0026$) \\ 
&  & friend, listed count & 89.904 ($\pm 0.0028$) \\ \midrule

\multirow{4}{*}{100} & \multirow{2}{*}{3} & friend, follower, listed count & 98.088 ($\pm 0.0037$) \\
&  & friend, follower, favorite count & 97.425 ($\pm 0.0058$) \\
& \multirow{2}{*}{2} & friend, follower count & 97.099 ($\pm 0.0051$) \\
&  & friend, listed count & 95.938 ($\pm 0.0073$) \\ \midrule

\multirow{4}{*}{10} & \multirow{2}{*}{3} & friend, follower, favorite count & 99.790 ($\pm 0.0014$) \\
&  & friend, favorites, listed count & 99.722 ($\pm 0.0015$) \\
& \multirow{2}{*}{2} & friend, favorites count & 99.639 ($\pm 0.0016$) \\
&  & follower, friend count & 99.483 ($\pm 0.0022$) \\

\bottomrule
\bottomrule

\end{tabular}%
}
\end{table}

\begin{table}[t]
\caption{RF Classification using dynamic features.}
\label{NAC-RF}
\centering
\resizebox{.9\columnwidth}{!}{%

\begin{tabular}{cclc} 
\toprule
{$u$} & {$n$} & {features} & {accuracy} \\ \midrule

\multirow{4}{*}{10,000} & \multirow{2}{*}{3} & friend, follower, favorite count & 94.408 ($\pm 0.0008$) \\
&  & friend, follower, status count & 94.216 ($\pm 0.0006$) \\
& \multirow{2}{*}{2} & friend, follower count & 81.105 ($\pm 0.0005$) \\
&  & friend, favorite count & 75.704 ($\pm 0.0006$) \\ \midrule

\multirow{4}{*}{1,000} & \multirow{2}{*}{3} & friend, follower, favorite count & 96.982 ($\pm 0.0008$) \\
&  & friend, follower, status count & 96.701 ($\pm 0.0008$) \\
& \multirow{2}{*}{2} & friend, follower count & 90.889 ($\pm 0.003$) \\ 
&  & friend, favorite count & 89.271 ($\pm 0.004$) \\ \midrule

\multirow{4}{*}{100} & \multirow{2}{*}{3} & friend, follower, favorite count & 99.286 ($\pm 0.0014$) \\
&  & friend, listed, favorite count & 99.149 ($\pm 0.0017$) \\
& \multirow{2}{*}{2} & friend, follower count & 97.690 ($\pm 0.0029$) \\
&  & listed, friend count & 97.275 ($\pm 0.00363$) \\ \midrule

\multirow{4}{*}{10} & \multirow{2}{*}{3} & friend, listed, favorite count & 99.942 ($\pm 0.0005$) \\
&  & follower, favorites, friend count & 99.930 ($\pm 0.00061$) \\
& \multirow{2}{*}{2} & friend, listed count & 99.885 ($\pm 0.0008$) \\
&  & follower, friend count & 99.776 ($\pm 0.0013$) \\

\bottomrule
\bottomrule

\end{tabular}%
}
\end{table}

By dynamic attributes we mean all those attributes that are likely to change over time. From Table \ref{input_description} these are the counts for: friends, followers, lists, statuses, and favorites, as well as a categorical representation of the time stamp based on the hour of each post.

Table \ref{NAC-KNN} presents the two best performing combinations in terms of accuracy for each of the values we consider for $n$ and $u$. In 10,000 users there is a 92\% chance of finding the correct account given the number of friends, followers, and the number of times an account has been listed. 

Table \ref{NAC-RF} presents similar results for the RF algorithm. Even without the ACT, we are able to achieve 94.41\% accuracy in a group of 10,000 users. These results are directly linked to the behavior of an account and are obtained from a multi-class model.

\subsubsection{Combining Static and Dynamic Attributes}
\label{completeAnalysis}

\begin{table}[t]
\caption{Accuracy of the top combination for \textit{n} number of inputs for the KNN classifier.}
\label{top_KNN}

\centering
\resizebox{0.9\columnwidth}{!}{%

\begin{tabular}{cclS} 
\toprule
{$u$} & {$n$} & {features} & {accuracy(\%)} \\ \midrule

\multirow{3}{*}{10,000} 
& {3} & day, minute, second & 96.737 $(\pm 0.019)$ \\
& {2} & follower, friend count & 83.719 $(\pm 0.021)$ \\
& {1} & friend count & 14.612 $(\pm 0.036)$ \\ \midrule

\multirow{3}{*}{1,000} 
& {3} & listed count, minute, second & 99.648 $(\pm 0.022)$ \\
& {2} & friend, listed count & 92.809 $(\pm 0.050)$ \\
& {1} & friend count & 40.151 $(\pm 0.089)$ \\ \midrule

\multirow{3}{*}{100} 
& {3} & month, minute, second & 100.00 $(\pm 0.000)$ \\
& {2} & minute, second & 98.836 $(\pm 0.101)$ \\
& {1} & friend count & 78.650 $(\pm 0.330)$ \\ \midrule

\multirow{3}{*}{10} 
& {3} & month, minute, second & 100.00 $(\pm 0.000)$ \\
& {2} & month, minute & 100.00 $(\pm 0.000)$ \\
& {1} & friend count & 96.428 $(\pm 0.312)$\\ 

\bottomrule
\bottomrule

\end{tabular}%
}
\end{table}

\begin{table}[t]
\centering
\caption{Accuracy of the top combination for \textit{n} number of inputs for the RF classifier.}
\resizebox{0.9\columnwidth}{!}{%
\begin{tabular}{cclS} 
\toprule
{$u$} & {$n$} & {features} & {accuracy(\%)} \\ \midrule

\multirow{3}{*}{10,000} 
& {3} & listed count, day, second & 94.234 $(\pm 0.022)$ \\
& {2} & friend count, minute & 81.352 $(\pm 0.347 )$ \\
& {1} & friend count & 23.958 $(\pm 0.089)$ \\ \midrule

\multirow{3}{*}{1,000} 
& {3} & friend count, minute, second & 99.881 $(\pm 0.008)$\\
& {2} & friend count, second & 97.28 $(\pm 0.032)$ \\
& {1} & friend count & 49.538 $(\pm 0.086)$ \\ \midrule

\multirow{3}{*}{100} 
& {3} & day, minute, second & 100.00 $(\pm 0.000)$ \\
& {2} & friend count, minute & 99.595 $(\pm 0.023)$ \\
& {1} & friend count & 81.489 $(\pm 0.299)$ \\ \midrule

\multirow{3}{*}{10} 
& {3} & day, minute, second & 100.00 $(\pm 0.000)$ \\
& {2} & day, second & 100.00 $(\pm 0.000)$ \\
& {1} & friend count & 96.858 $(\pm 0.256)$\\ 

\bottomrule
\bottomrule

\end{tabular}%
}
\label{top_RF}
\end{table}
As we expected from our feature selection, the top combinations per value of $n$ inputs are different per classifier. Tables \ref{top_KNN}, \ref{top_RF}, \ref{top_MLR} show the accuracy and the error obtained for the best performing pair of features aggregated over all runs for the three classifiers. We can see that the least accurate predictions are those derived by means of the MLR algorithm. Then, probably for its robustness against noise, RF performs best for the smaller user-groups, but it is KNN that provides the best performance for the case where $u=10,000$.

\begin{table}[t]
\caption{Accuracy  of the top combination for \textit{n} number of inputs for the MLR classifier.}
\label{top_MLR}

\centering
\resizebox{0.9\columnwidth}{!}{%

\begin{tabular}{cclS} 
\toprule
{$u$} & {$n$} & {features} & {accuracy(\%)} \\ \midrule

\multirow{3}{*}{10,000} 
& {3} & day, minute, second & 96.060 $(\pm 0.060)$ \\
& {2} & day, minute & 29.571 $(\pm 5.11 )$ \\
& {1} & second & 2.241 $(\pm 0.080)$ \\ \midrule

\multirow{3}{*}{1,000} 
& {3} & hour, minute, second & 99.329 $(\pm 0.060)$\\
& {2} & day, minute & 61.77 $(\pm 5.11)$ \\
& {1} & hour & 3.105 $(\pm 0.080)$ \\ \midrule

\multirow{3}{*}{100} 
& {3} & day, minute, second & 100.00 $(\pm 0.000)$ \\
& {2} & second, minute & 98.494 $(\pm 0.116)$ \\
& {1} & second & 26.303  $(\pm 0.536)$ \\ \midrule

\multirow{3}{*}{10} 
& {3} & day, minute, second & 100.00 $(\pm 0.000)$ \\
& {2} & day, minute & 100.00 $(\pm 0.000)$ \\
& {1} & second & 94.129 $(\pm 0.702)$\\ 

\bottomrule
\bottomrule

\end{tabular}%
}
\end{table}

In general, the classification task gets incrementally more challenging as the number of users increases. We are able to achieve a 90\% accuracy over all the classifiers with respect to a 0.01\% baseline offered by the random case. If we consider the 10 most likely output candidates (i.e., the top-10 classes), for the 10,000 user group there is a 99.22\% probability of finding the user. 
\begin{figure*}[t]
\minipage{0.32\textwidth}
  \includegraphics[width=\linewidth]{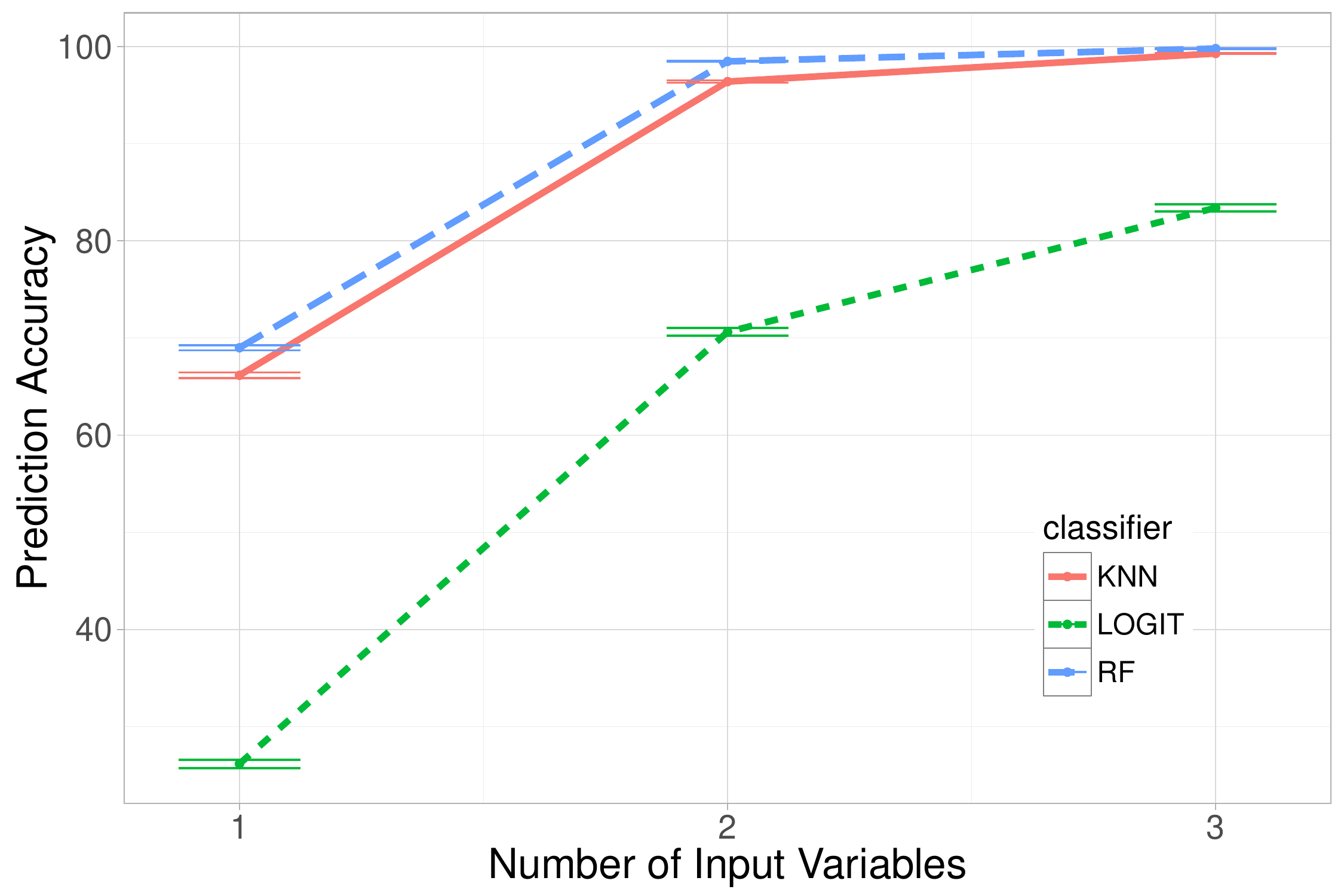}
  \caption{Overall accuracy $u=100$.}
  \label{ug_100}
\endminipage\hfill
\minipage{0.32\textwidth}%
  \includegraphics[width=\linewidth]{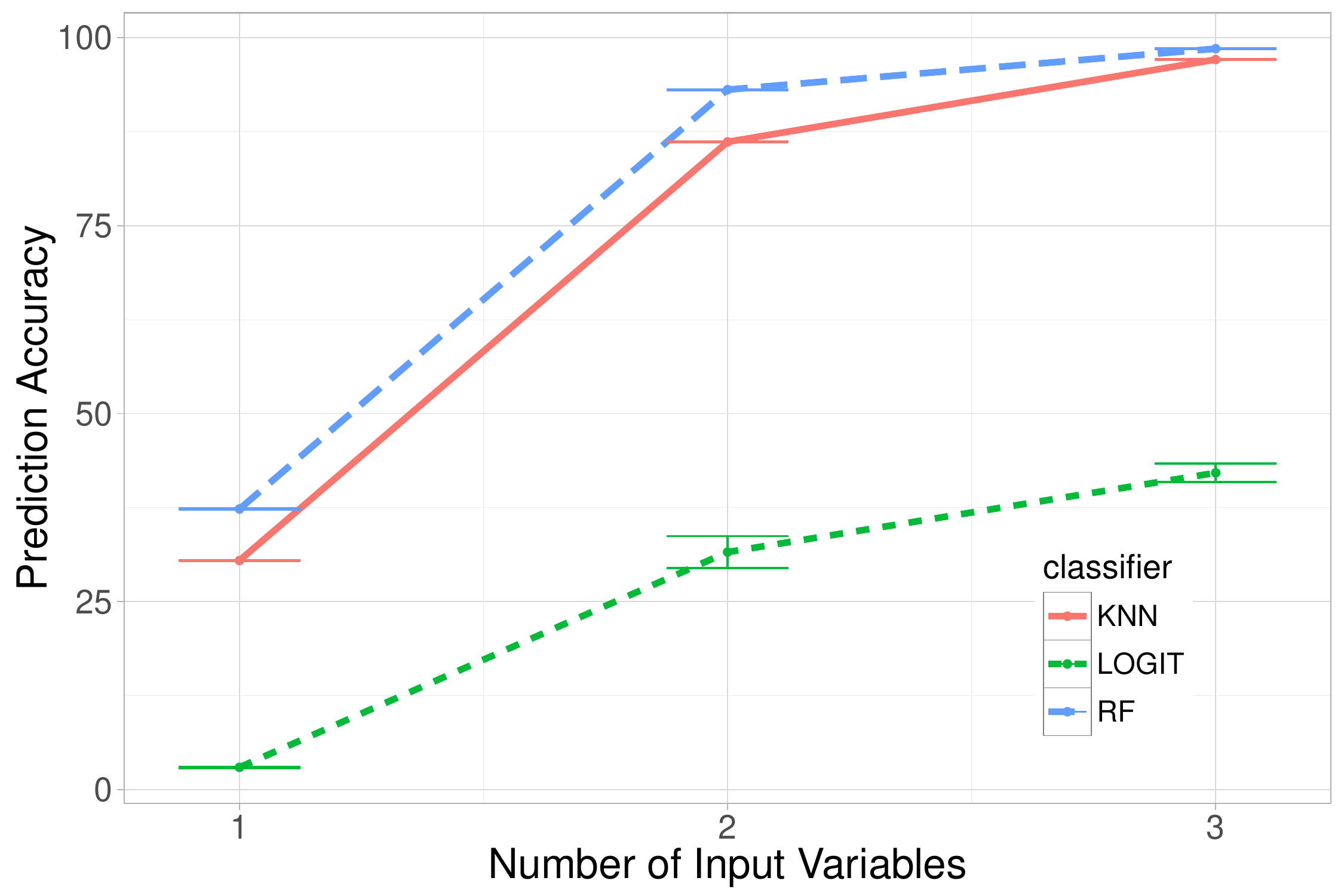}
  \caption{Overall accuracy $u=1,000$.}
  \label{ug_1k}
\endminipage
\minipage{0.32\textwidth}%
  \includegraphics[width=\linewidth]{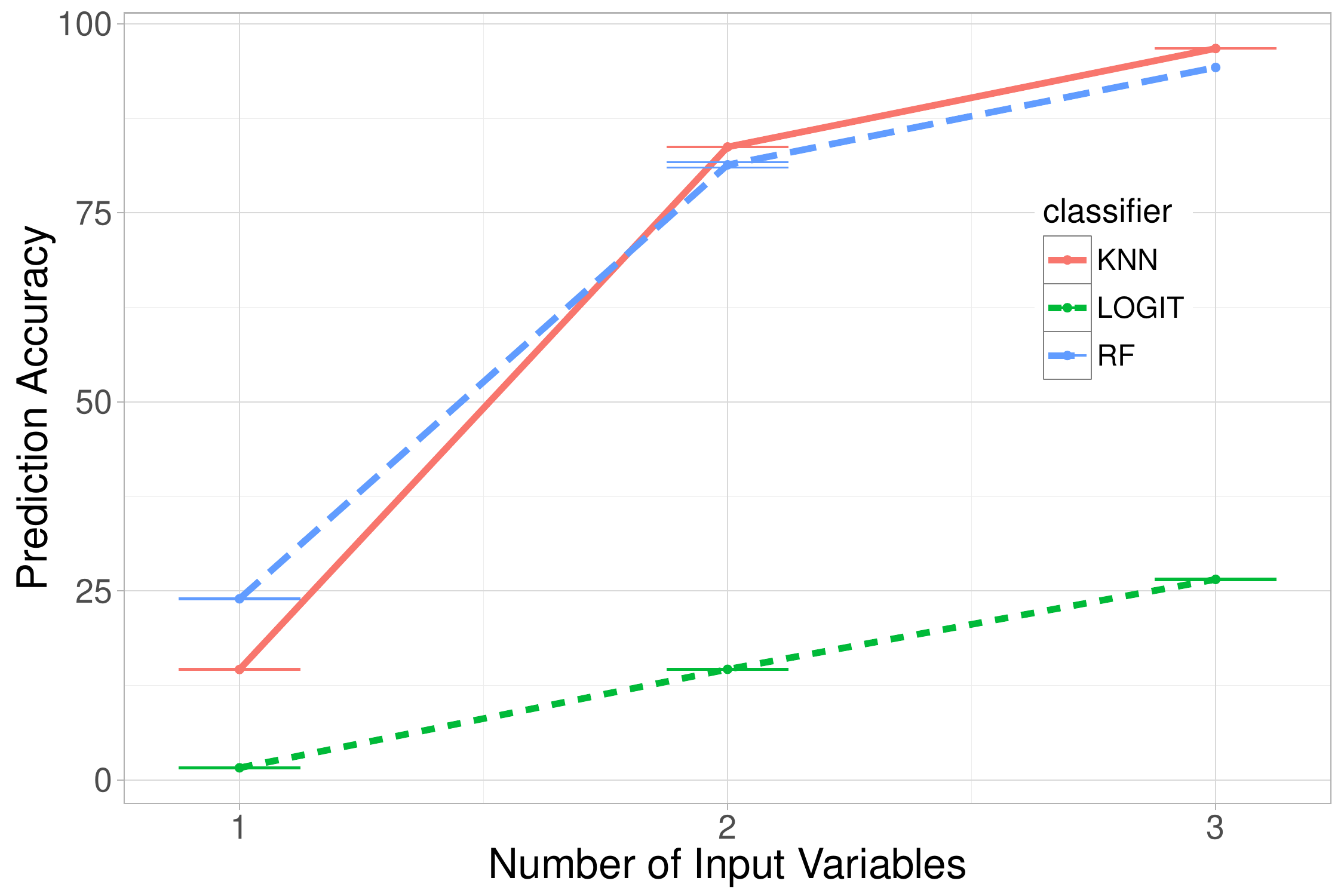}
  \caption{Overall accuracy $u=10,000$.}
  \label{ug_10k}
\endminipage
\end{figure*}

\begin{figure*}[t]
\centering
\minipage{0.32\textwidth}
  \includegraphics[width=\linewidth]{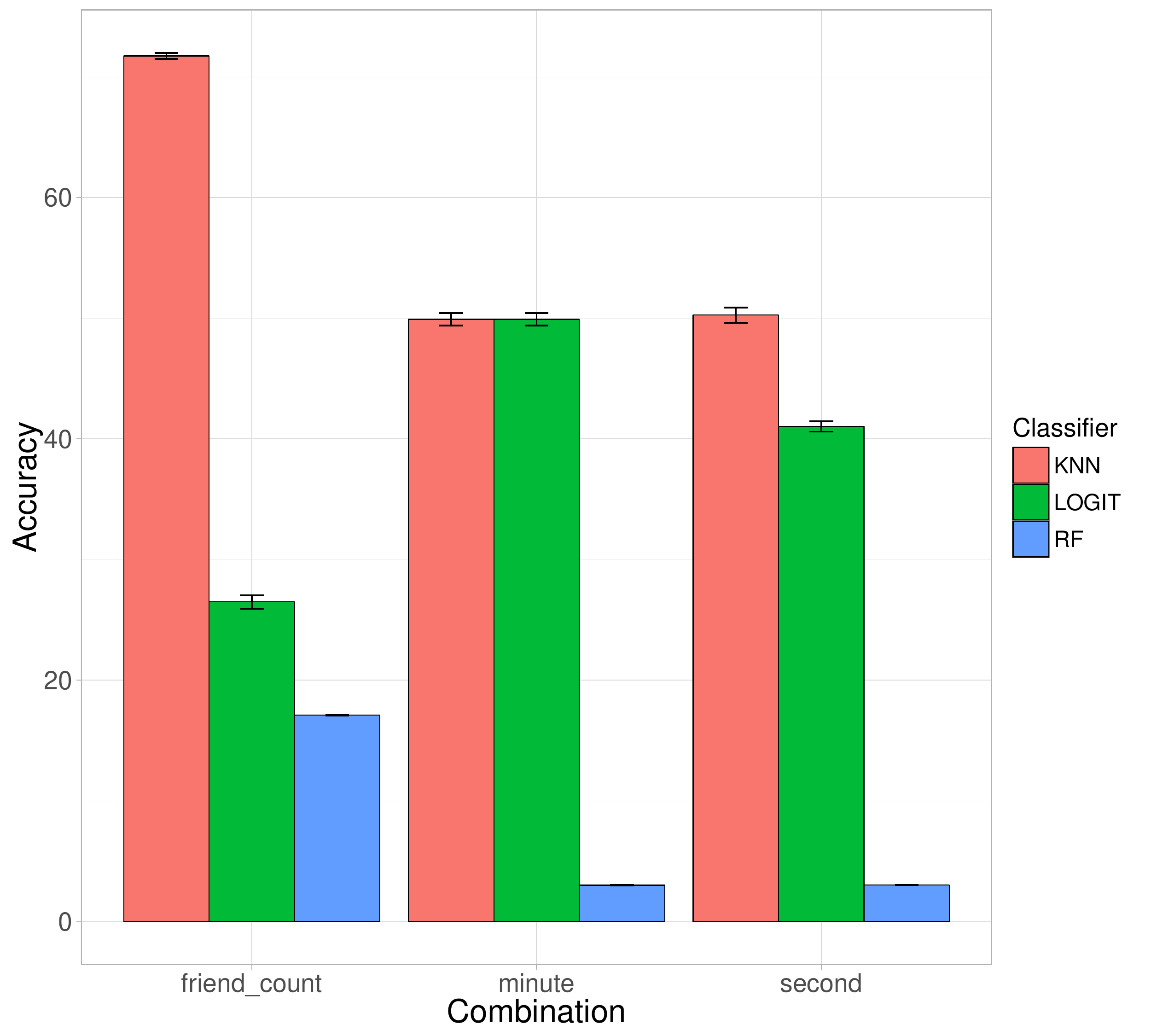}
  \caption{Performance of the most popular features for \textit{n}=1.}
  \label{popularity_n1}
\endminipage\hfill
\minipage{0.32\textwidth}
  \includegraphics[width=\linewidth]{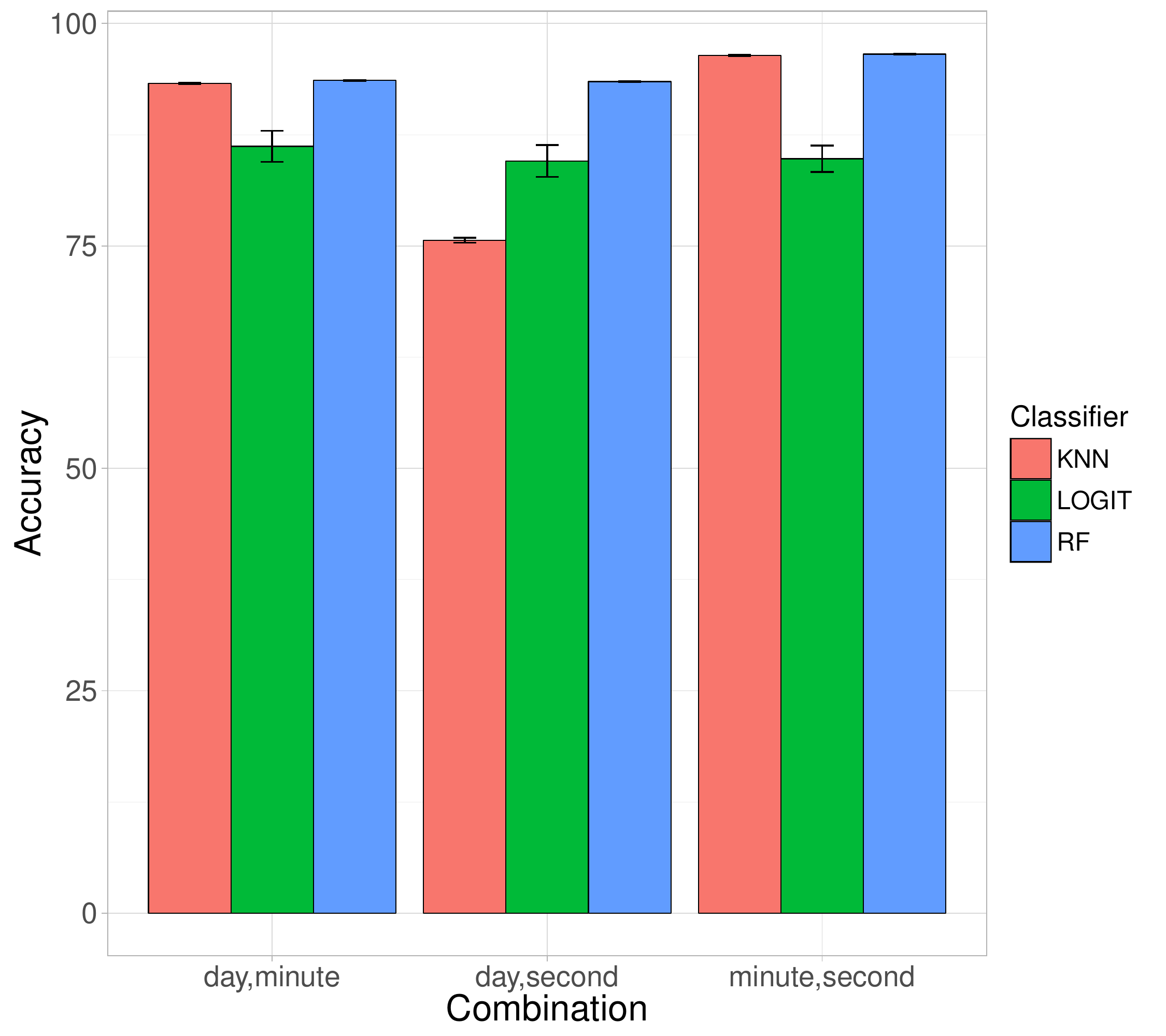}
  \caption{Performance of the most popular features for \textit{n}=2.}
  \label{popularity_n2}
\endminipage\hfill
\minipage{0.32\textwidth}%
  \includegraphics[width=\linewidth]{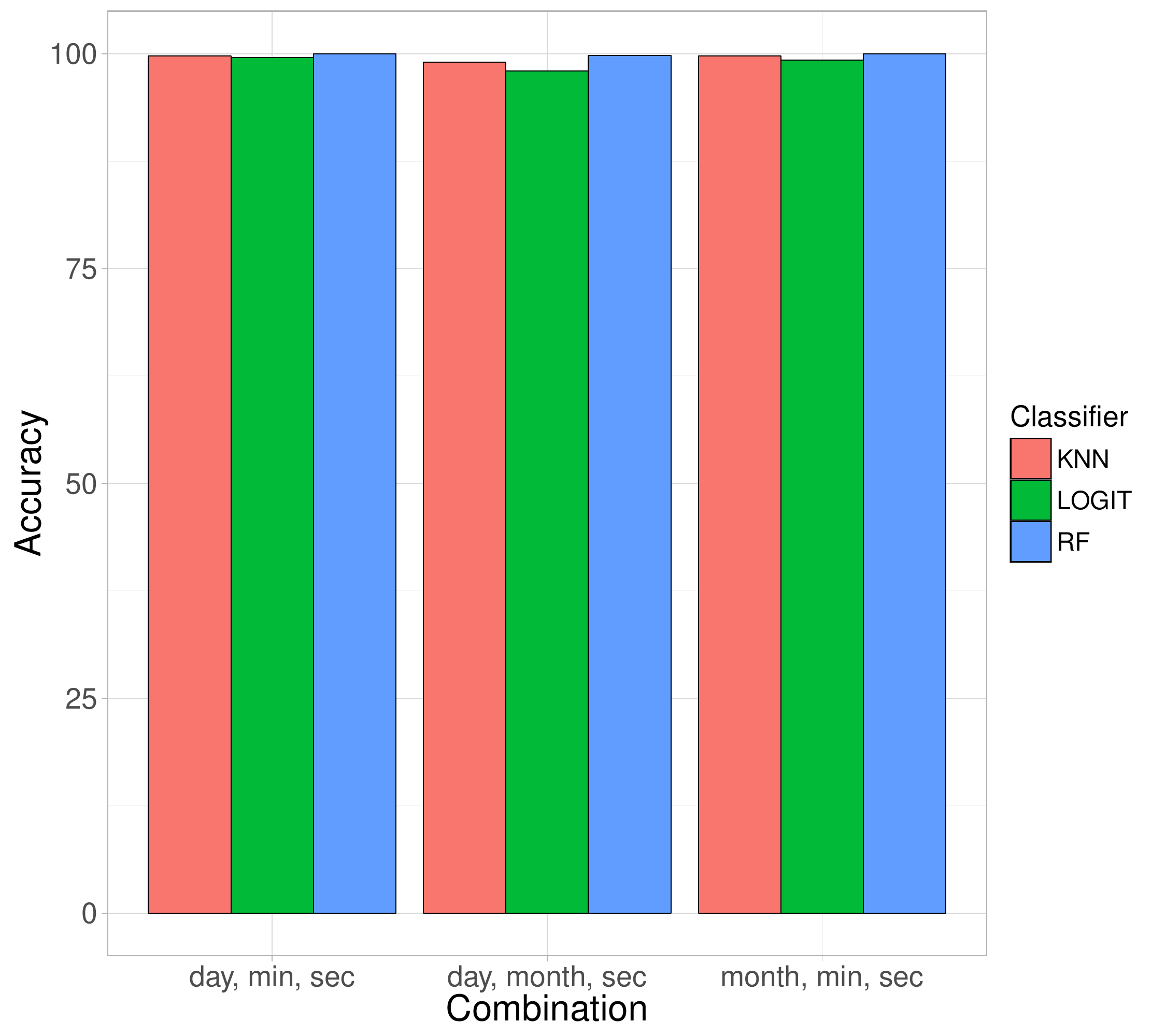}
  \caption{Performance of the most popular features for \textit{n}=3.}
  \label{popularity_n3}
\endminipage%
\end{figure*}

Finally, we looked at how the best performing combinations across all algorithms performed in models for each classifier. 
The top three combinations per value of $n$ are presented in Figures \ref{popularity_n1}, \ref{popularity_n2}, \ref{popularity_n3}. For the same value of $u$ as we increase $n$ the accuracy increases.

\subsection{Obfuscation}
\begin{figure*}[t]
\centering
\minipage{0.32\textwidth}
  \includegraphics[width=\linewidth]{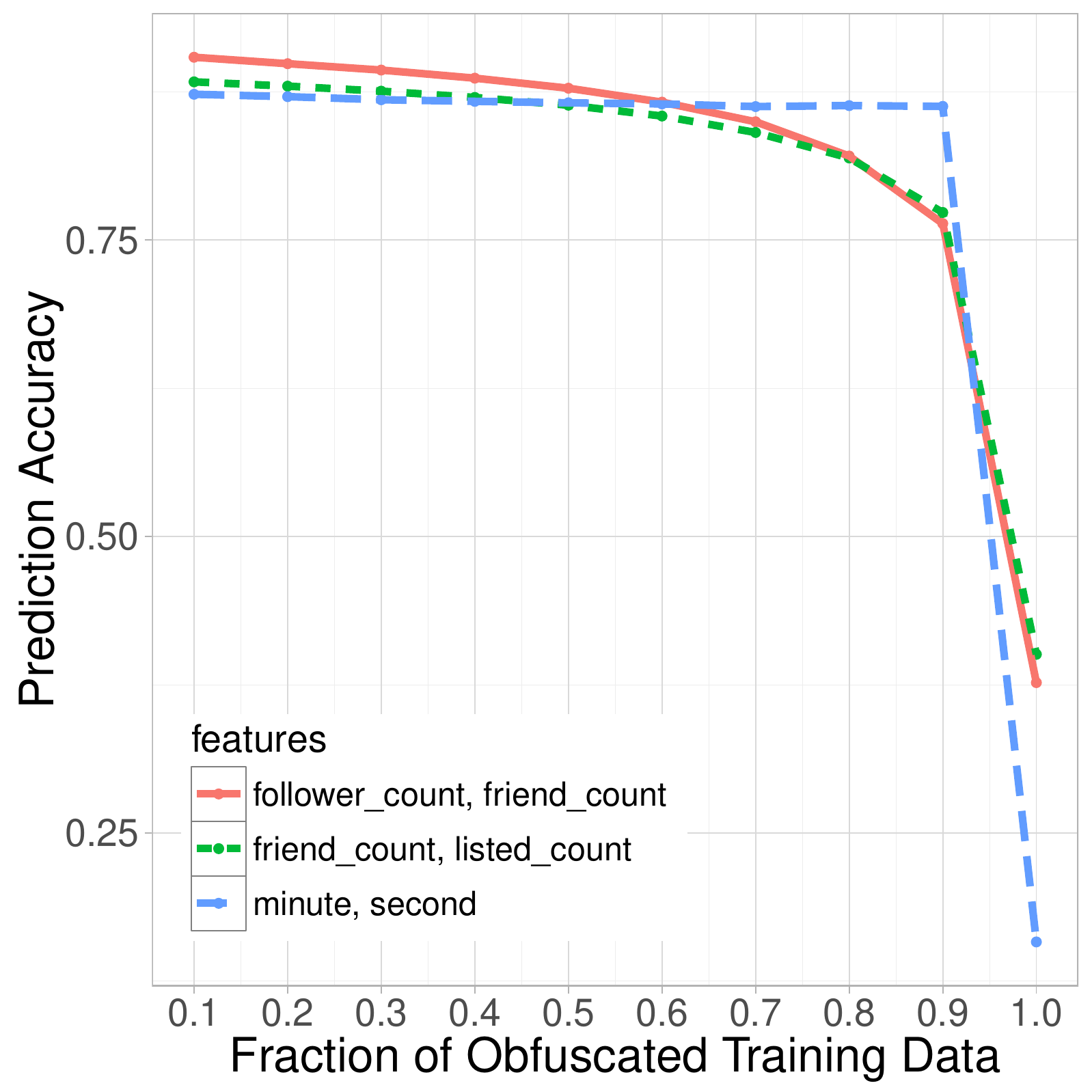}
  \caption{KNN Change in predictive accuracy with obfuscated training data.}
  \label{knn_obfs}
\endminipage\hfill
\minipage{0.32\textwidth}
  \includegraphics[width=\linewidth]{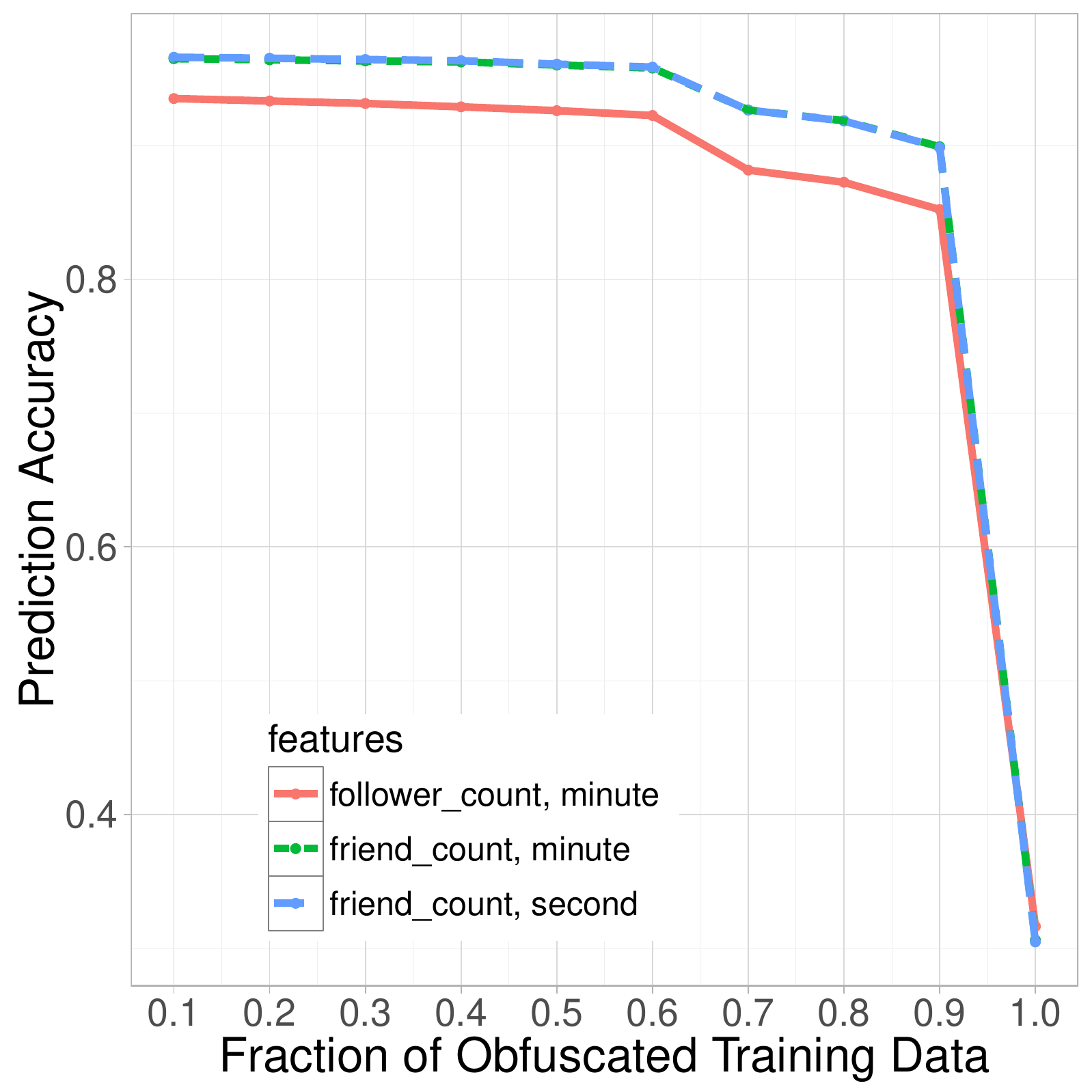}
  \caption{RF Change in predictive accuracy with obfuscated training data.}
  \label{rf_obfs}
\endminipage\hfill
\minipage{0.32\textwidth}%
  \includegraphics[width=\linewidth]{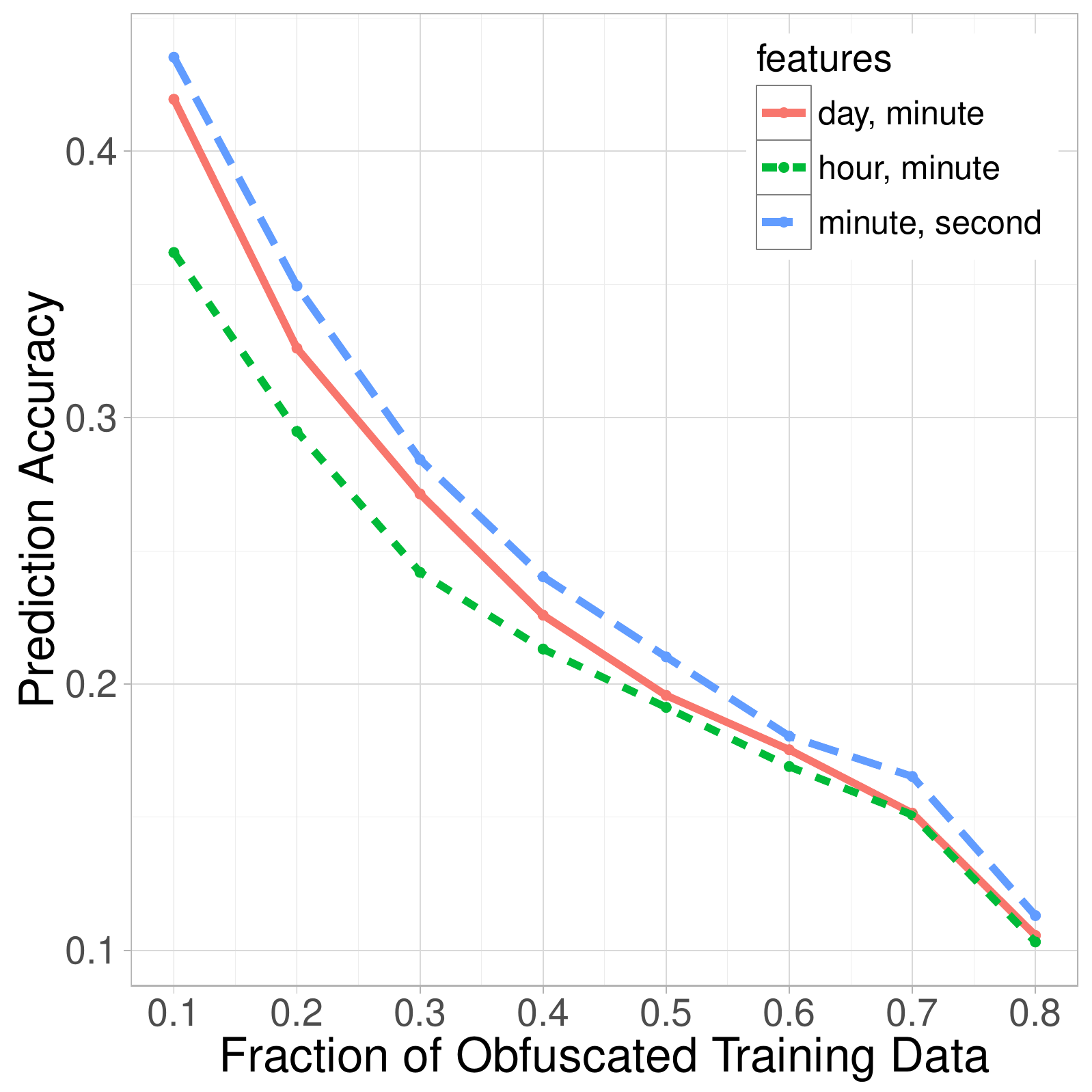}
  \caption{MLR Change in predictive accuracy with obfuscated training data.}
  \label{mlr_obfs}
\endminipage%
\end{figure*}
The final analysis looks at the effects of data anonymization and data randomization techniques on the proposed methodology. 
As we state in
the Method section
we start with the original data set then apply a rounding algorithm to change the values of each reading. The number of readings given in input to the algorithm increase in steps of 10\% from no anonymization to 100\% perturbation where we show full randomization. 
To test obfuscation, we selected the 3 most accurate combinations of features for $n=2$ and $u=1,000$ for each of the classification methods. 

While we are not working with geospatial data, we find that similar to~\cite{laplace}, the level of protection awarded by perturbation is not very significant until we get to 100\% randomization. Figures \ref{knn_obfs}, \ref{rf_obfs}, and \ref{mlr_obfs} show how each algorithm performs with an increasing number of obfuscated points. RF is the best performing providing the most accurate result despite data anonymity. MLR is the most sensitive of the three. Even with 20\% randomization there is a steep decrease in terms of prediction accuracy. 

\subsection{Execution Time}
\label{executionTime}
To compare the performance of the classifiers in terms of execution time we used a dedicated server with eight core Intel Xeon E5-2630 processors with 192GB DDR4 RAM running at 2,133MHz. For the implementation of the algorithms, we used Python 2.7 and Sci-kit learn release 0.17.1. 

\begin{figure}[t]
\caption{Mean execution time as a function of features.}
\label{timing_2}
\centering
\resizebox{\columnwidth}{!}{%
\includegraphics[width=\linewidth]{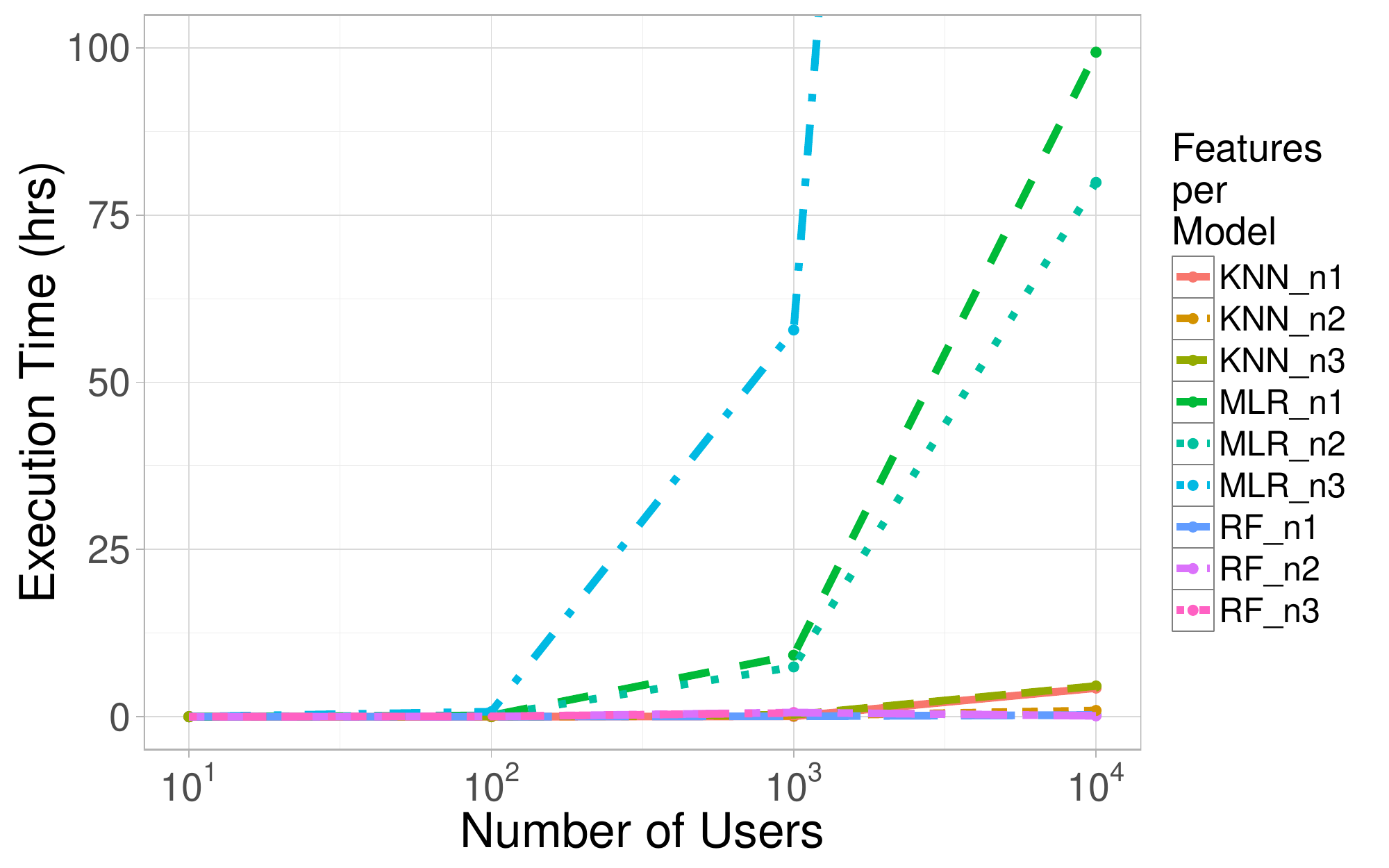}%
}
\end{figure}

\begin{figure}[t]
\caption{F-score for increasing intermediate sample sizes.}
\label{dAc}
\centering
\resizebox{\columnwidth}{!}{%
\includegraphics[width=\linewidth]{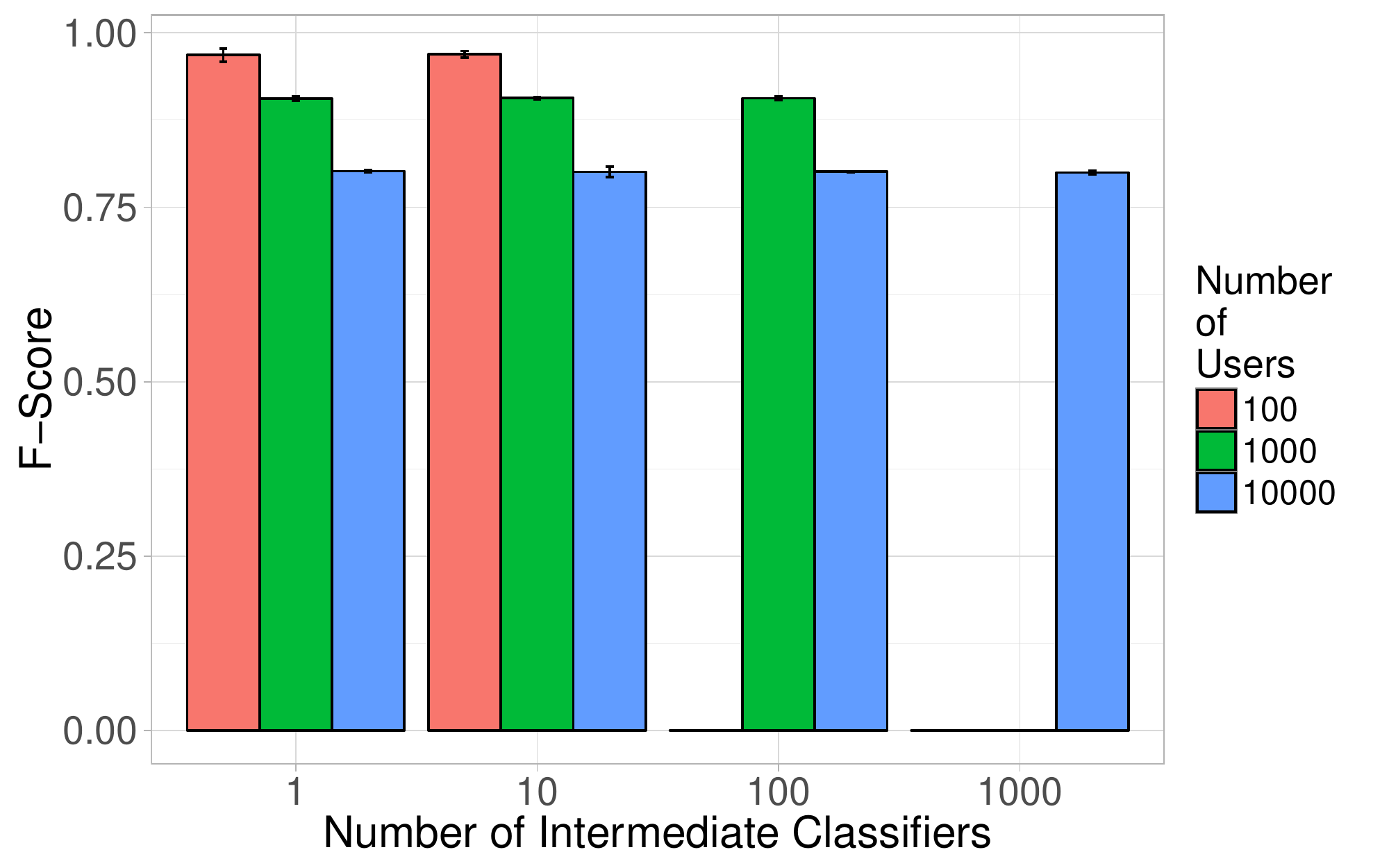}%
}
\end{figure}

Figure \ref{timing_2} shows execution time as a function of the number of output classes in each model. Note that the performance gap between MLR and the other two is significant. While KNN and RF show a linear increase over the number of users, the rate of change for MLR is much more rapid. At $u=1,000$ and $n=3$, for example, MLR is 105 times slower than RF and 210 times slower than KNN. The performance bottleneck for multi-class classifiers is the number of output classes. Finding a viable solution is fundamental for this project and for the general applicability of the method. 

To address this, we implemented a \textit{divide and conquer} algorithm where we get intermediate predictions over smaller subsets of classes and build one final model with the (intermediate) results. This method allows for faster execution and lower memory requirements for each model and parallel execution resulting in further performance enhancements. 
Figure \ref{dAc} shows the effectiveness of the proposed method for $u=100, 1,000,$ and $10,000$. 
We also observe that the accuracy is independent from the number of subsets.

\section{Discussion and Limitations}

In this paper we have presented a comparison of three classification methods in terms of accuracy and execution time. We reported their performance with and without input obfuscation. Overall, KNN provides the best trade-off in terms of accuracy and execution time with both the original and the obfuscated data sets. 
We tested each of the algorithms following the wrapper method by increasing the number of input features in each model in steps of one. The results, summarized in Tables \ref{top_KNN}, \ref{top_RF}, and \ref{top_MLR}, show that the metadata can be effectively exploited to classify individuals inside a group of 10,000 randomly selected Twitter users. While both KNN and RF are similar in terms of accuracy and execution, MLR is consistently outperformed. Moreover, as shown in Figures \ref{ug_100}, \ref{ug_1k}, and \ref{ug_10k}, as the number of output classes increases, the difference in performance becomes more pronounced. 
It is also important to note that, as shown in Table \ref{top_MLR}, the accuracy exhibited by MLR depends entirely on the six constituent features of the account creation time. Finally, the performance of MLR is the most sensitive to obfuscation. As shown in Figure \ref{mlr_obfs}, the rounding algorithm results in a monotonic drop in accuracy for the best performing combinations. 

One challenge in multi-class classification lies in the scalability of the algorithm. We found that while both the number of input features and the number of output classes have a detrimental impact on performance, the bottleneck of the algorithm is caused by the number of output classes. To address this, we implemented a \textit{divide and conquer} algorithm that partitions the users and creates models with fewer output classes. The intermediate outputs of each of the smaller models are then combined to produce a single result. Figure \ref{dAc} shows that this implementation does not affect the accuracy of the classification algorithms. As shown in the figure, varying the size of the partitions results in the same precision and recall for the same total number of users as compared to the results obtained from a single model. Having shown that the results are equivalent the implementation becomes a contribution from our work: partitioning the number of output classes in intermediate steps results in less memory consumption and faster execution times.

We now discuss the implications of our findings on devising techniques for data privacy and anonymization. Previous studies have proposed numerous techniques for privacy in social-media data mining by focusing on the content of data rather than its metadata ~\cite{continuousAuth,rao2010classifying,IndividualEmotions,hometown}. In this study we have demonstrated that the metadata could also play a significant role in revealing user identities. Our findings demonstrate that generic information such as the number of friends, the number of favorited tweets, the number of followers is sufficient to distinguish a user from another with an accuracy of over 90\%. 
These results have implications for researchers and practitioners that share datasets: great care has to be taken not only in obfuscating identities in primary data (such as posts and profile information) but also in the metadata (auxiliary fields) associated to them. More specifically, special attention has to be devoted to the fact that combinations of input features reveal user's identities as was shown in this work.

\section{Related Work}
%
\subsection{Privacy}
``Privacy is measured by the information gain of an observer"~\cite{tclose}. If an attacker is able to either trace an entry to an account, or an account to a user, that attacker gains information and the user loses of privacy.
With regards to online social networks, the risks to privacy presented in~\cite{alterman2003} are magnified when considering the temporal dimension. In contrast, the privacy controls that have been introduced are incomplete. Studies have shown that often times these controls are lacking in terms of safeguarding against a dedicated adversary~\cite{deanonymizing} but, most importantly, they are difficult to use and are yet to be adopted by the majority of users~\cite{privacySuites}. This results in a wealth of information about users openly available on the Web. Primary personal data might be anonymized, but little attention has been devoted to metadata. This paper shows the privacy risks associated to metadata in terms of user identification, demonstrating that obfuscation is often insufficient to protect users or the trade-off between obfuscation and utility of the data might not be acceptable.

\subsection{Identification of Individuals through Metadata}
Various identification methods have been developed to exploit existing information about physical attributes, behavior and choices of a user. 
For example, in ~\cite{continuousAuth} the authors describe an active authentication system that uses both physical and behavioral attributes to continuously monitor the identity of the user. In~\cite{typing} the authors give an overview and present the advantages and disadvantages of keystroke characterization for authentication, identification, and monitoring. Finally, in~\cite{frey2016lightweight} the authors use smartphone metadata, namely the apps that have been installed on a phone, to create a signature achieving a 99.75\% accuracy. In this work, for the first time we present an in-depth analysis of the risks associated to online social networks metadata. We would like to stress that metadata is more easily available and/or accessible and, for this reason, it represents a significant risk for user privacy.

\subsection{Identification of Devices through Metadata}
In~\cite{phoneDiagnostics} the authors use phone diagnostics, i.e., information such as hardware statistics and system settings, to uniquely identify devices. Instead, the method presented in~\cite{remotedf} relies on small but measurable differences of a device's hardware clock. In~\cite{bojinov2014mobile} the authors exploit some characteristics of hardware components, in this case the accelerometer, the microphone and the speaker to create a fingerprint without consent or knowledge of users. This is information about the sensors and not the measurements collected with them: for this reason, we characterize each of these as metadata.

\section{Conclusions}

In this paper, we have used Twitter as a case study to quantify the uniqueness of the association between metadata and user identity, devising techniques for user identification and related obfuscation strategies. We have tested the performance of three state-of-the-art machine learning algorithms, MLR, KNN and RF using a corpus of 5 million Twitter users.  KNN provides the best performance in terms of accuracy for an increasing number of users and obfuscated data. We demonstrated that through this algorithm, we are able to identify 1 user in a group of 10,000 with approximately 96.7\% accuracy.  Moreover, if we broaden the scope of our search and consider the best 10 candidates from a group of 10,000 users, we achieve a 99.22\% accuracy. We also demonstrated that obfuscation strategies are ineffective: after perturbing 60\% of the training data, it is possible to classify users with an accuracy greater than 95\%.

We believe that this work will contribute to raising awareness of the privacy risks associated to metadata. It is worth underlining that, even if we focused on Twitter for the experimental evaluation, the methods described in this work can be applied to a vast class of platforms and systems that generate metadata with similar characteristics. This problem is particularly relevant given the increasing number of organizations that release open data with metadata associated to it or the popularity of social platforms that offer APIs to access their data, which is often accompanied by metadata. 

\section{Acknowledgments}
This work was supported by The Alan Turing Institute under the EPSRC grant EP/N510129/1 and at UCL through the EPSRC grant EP/P016278/1.

\begin{quote}
\begin{small}
\bibliographystyle{aaai}
\bibliography{bib_2}

\begin{thebibliography}{}

\bibitem[\protect\citeauthoryear{Agrawal and Srikant}{}]{ppdmAgrawal}
Agrawal, R., and Srikant, R.
\newblock {Privacy-Preserving Data Mining}.
\newblock In {\em SIGMOD'00}.

\bibitem[\protect\citeauthoryear{Alterman}{2003}]{alterman2003}
Alterman, A.
\newblock 2003.
\newblock {A piece of yourself: Ethical issues in biometric identification}.
\newblock {\em Ethics and Information Technology} 5(3):139--150.

\bibitem[\protect\citeauthoryear{Bailey, Okolica, and
  Peterson}{2014}]{Bailey201477}
Bailey, K.~O.; Okolica, J.~S.; and Peterson, G.~L.
\newblock 2014.
\newblock {User identification and authentication using multi-modal behavioral
  biometrics }.
\newblock {\em Computers \& Security} 43:77 -- 89.

\bibitem[\protect\citeauthoryear{Bakken \bgroup et al\mbox.\egroup
  }{2004}]{desensitization}
Bakken, D.~E.; Rarameswaran, R.; Blough, D.~M.; Franz, A.~A.; and Palmer, T.~J.
\newblock 2004.
\newblock {Data obfuscation: anonymity and desensitization of usable data
  sets}.
\newblock {\em IEEE Security and Privacy} 2(6):34--41.

\bibitem[\protect\citeauthoryear{Benevenuto \bgroup et al\mbox.\egroup
  }{}]{benevenuto2010detecting}
Benevenuto, F.; Magno, G.; Rodrigues, T.; and Almeida, V.
\newblock {Detecting Spammers on Twitter}.
\newblock In {\em CEAS'10}.

\bibitem[\protect\citeauthoryear{Bishop}{2001}]{bishop}
Bishop, C.~M.
\newblock 2001.
\newblock {\em Pattern Recognition and Machine Learning}.
\newblock Springer, New York.

\bibitem[\protect\citeauthoryear{Bo \bgroup et al\mbox.\egroup
  }{}]{SilentSense}
Bo, C.; Zhang, L.; Li, X.-Y.; Huang, Q.; and Wang, Y.
\newblock Silentsense: Silent user identification via touch and movement
  behavioral biometrics.
\newblock In {\em MobiCom'13}.

\bibitem[\protect\citeauthoryear{Bojinov \bgroup et al\mbox.\egroup
  }{2014}]{bojinov2014mobile}
Bojinov, H.; Michalevsky, Y.; Nakibly, G.; and Boneh, D.
\newblock 2014.
\newblock Mobile device identification via sensor fingerprinting.
\newblock {\em arXiv preprint arXiv:1408.1416}.

\bibitem[\protect\citeauthoryear{Bollen, Mao, and Pepe}{}]{groupEmotions}
Bollen, J.; Mao, H.; and Pepe, A.
\newblock Modeling public mood and emotion: Twitter sentiment and
  socio-economic phenomena.
\newblock In {\em ICWSM'11}.

\bibitem[\protect\citeauthoryear{Bonneau, Anderson, and
  Church}{}]{privacySuites}
Bonneau, J.; Anderson, J.; and Church, L.
\newblock Privacy suites: shared privacy for social networks.
\newblock In {\em SOUPS'09}.

\bibitem[\protect\citeauthoryear{Breiman}{2001}]{randomForest}
Breiman, L.
\newblock 2001.
\newblock {Random Forests}.
\newblock {\em Machine Learning} 45(1):5--32.

\bibitem[\protect\citeauthoryear{De~Cristofaro \bgroup et al\mbox.\egroup
  }{}]{hummingbird}
De~Cristofaro, E.; Soriente, C.; Tsudik, G.; and Williams, A.
\newblock {Hummingbird: Privacy at the time of Twitter}.
\newblock In {\em SP'12}.

\bibitem[\protect\citeauthoryear{Frey, Xu, and Ilic}{}]{frey2016lightweight}
Frey, R.~M.; Xu, R.; and Ilic, A.
\newblock A lightweight user tracking method for app providers.
\newblock In {\em CF'16}.

\bibitem[\protect\citeauthoryear{Hays and Efros}{}]{im2gps}
Hays, J., and Efros, A.
\newblock Im2gps: estimating geographic information from a single image.
\newblock In {\em CVPR'08}.

\bibitem[\protect\citeauthoryear{Huang, Yang, and Chuang}{2008}]{classifier}
Huang, C.-J.; Yang, D.-X.; and Chuang, Y.-T.
\newblock 2008.
\newblock {Application of wrapper approach and composite classifier to the
  stock trend prediction}.
\newblock {\em Expert Systems with Applications} 34(4):2870--2878.

\bibitem[\protect\citeauthoryear{Humphreys, Gill, and
  Krishnamurthy}{}]{humphreys2010much}
Humphreys, L.; Gill, P.; and Krishnamurthy, B.
\newblock {How much is too much? Privacy issues on Twitter}.
\newblock In {\em ICA'10}.

\bibitem[\protect\citeauthoryear{Jahanbakhsh, King, and Shoja}{2012}]{hometown}
Jahanbakhsh, K.; King, V.; and Shoja, G.~C.
\newblock 2012.
\newblock {They Know Where You Live!}
\newblock {\em CoRR} abs/1202.3504.

\bibitem[\protect\citeauthoryear{Kohno, Broido, and Claffy}{2005}]{remotedf}
Kohno, T.; Broido, A.; and Claffy, K.~C.
\newblock 2005.
\newblock Remote physical device fingerprinting.
\newblock {\em IEEE Transactions on Dependable and Secure Computing}
  2(2):93--108.

\bibitem[\protect\citeauthoryear{Li, Li, and Venkatasubramanian}{}]{tclose}
Li, N.; Li, T.; and Venkatasubramanian, S.
\newblock {t-Closeness: Privacy Beyond k-Anonymity and l-Diversity}.
\newblock In {\em ICDE'07}.

\bibitem[\protect\citeauthoryear{Liu and Nocedal}{1989}]{Liu1989}
Liu, D.~C., and Nocedal, J.
\newblock 1989.
\newblock On the limited memory bfgs method for large scale optimization.
\newblock {\em Mathematical Programming} 45(1):503--528.

\bibitem[\protect\citeauthoryear{Liu and Yu}{2005}]{featureSelection}
Liu, H., and Yu, L.
\newblock 2005.
\newblock {Toward integrating feature selection algorithms for classification
  and clustering}.
\newblock {\em IEEE Transactions on Knowledge and Data Engineering}
  17(4):491--502.

\bibitem[\protect\citeauthoryear{Malik, Ghazi, and Ali}{}]{pdmMalik}
Malik, M.~B.; Ghazi, M.~A.; and Ali, R.
\newblock {Privacy Preserving Data Mining Techniques: Current Scenario and
  Future Prospects}.
\newblock In {\em ICCCT'12}.

\bibitem[\protect\citeauthoryear{Mooney and Duval}{1993}]{bootstrapping}
Mooney, C., and Duval, R.
\newblock 1993.
\newblock {\em {Bootstrapping: A nonparametric approach to statistical
  inference}}.
\newblock Sage.

\bibitem[\protect\citeauthoryear{Mowbray, Pearson, and Shen}{2012}]{cloud2012}
Mowbray, M.; Pearson, S.; and Shen, Y.
\newblock 2012.
\newblock {Enhancing privacy in cloud computing via policy-based obfuscation}.
\newblock {\em The Journal of Supercomputing} 61(2):267--291.

\bibitem[\protect\citeauthoryear{Narayanan and Shmatikov}{}]{deanonymizing}
Narayanan, A., and Shmatikov, V.
\newblock De-anonymizing social networks.
\newblock In {\em SP'09}.

\bibitem[\protect\citeauthoryear{Patel \bgroup et al\mbox.\egroup
  }{2016}]{continuousAuth}
Patel, V.~M.; Chellappa, R.; Chandra, D.; and Barbello, B.
\newblock 2016.
\newblock Continuous user authentication on mobile devices: Recent progress and
  remaining challenges.
\newblock {\em IEEE Signal Processing Magazine} 33(4):49--61.

\bibitem[\protect\citeauthoryear{Peacock, Ke, and Wilkerson}{2004}]{typing}
Peacock, A.; Ke, X.; and Wilkerson, M.
\newblock 2004.
\newblock Typing patterns: a key to user identification.
\newblock {\em IEEE Security and Privacy} 2(5):40--47.

\bibitem[\protect\citeauthoryear{Pedregosa \bgroup et al\mbox.\egroup
  }{2011}]{scikit}
Pedregosa, F.; Varoquaux, G.; Gramfort, A.; Michel, V.; Thirion, B.; Grisel,
  O.; Blondel, M.; Prettenhofer, P.; Weiss, R.; Dubourg, V.; et~al.
\newblock 2011.
\newblock {Scikit-learn: Machine learning in Python}.
\newblock {\em Journal of Machine Learning Research} 12(Oct):2825--2830.

\bibitem[\protect\citeauthoryear{Polat and Du}{}]{collaborative}
Polat, H., and Du, W.
\newblock {Privacy-Preserving Collaborative Filtering Using Randomized
  Perturbation Techniques}.
\newblock In {\em ICMD'03}.

\bibitem[\protect\citeauthoryear{Quan, Yin, and Guo}{}]{laplace}
Quan, D.; Yin, L.; and Guo, Y.
\newblock {Enhancing the Trajectory Privacy with Laplace Mechanism}.
\newblock In {\em Trustcom'15}.

\bibitem[\protect\citeauthoryear{Quattrone \bgroup et al\mbox.\egroup
  }{}]{phoneDiagnostics}
Quattrone, A.; Bhattacharya, T.; Kulik, L.; Tanin, E.; and Bailey, J.
\newblock Is this you?: identifying a mobile user using only diagnostic
  features.
\newblock In {\em MUM'14}.

\bibitem[\protect\citeauthoryear{Rao \bgroup et al\mbox.\egroup
  }{}]{rao2010classifying}
Rao, D.; Yarowsky, D.; Shreevats, A.; and Gupta, M.
\newblock Classifying latent user attributes in twitter.
\newblock In {\em SMUC'10}.

\bibitem[\protect\citeauthoryear{Rossi and Musolesi}{}]{check-in}
Rossi, L., and Musolesi, M.
\newblock {It's the way you check-in: identifying users in location-based
  social networks}.
\newblock In {\em COSN'14}.

\bibitem[\protect\citeauthoryear{Shoshitaishvili, Kruegel, and
  Vigna}{2015}]{shoshitaishvili2015portrait}
Shoshitaishvili, Y.; Kruegel, C.; and Vigna, G.
\newblock 2015.
\newblock Portrait of a privacy invasion.
\newblock {\em Proceedings on Privacy Enhancing Technologies} 2015(1):41--60.

\bibitem[\protect\citeauthoryear{Silva \bgroup et al\mbox.\egroup
  }{}]{eatANDdrink}
Silva, T.~H.; de~Melo, P. O. S.~V.; Almeida, J.~M.; Musolesi, M.; and Loureiro,
  A. A.~F.
\newblock {You are What you Eat (and Drink): Identifying Cultural Boundaries by
  Analyzing Food {\&} Drink Habits in Foursquare}.
\newblock In {\em ICWSM'14}.

\bibitem[\protect\citeauthoryear{Stringhini, Kruegel, and
  Vigna}{}]{stringhini2010detecting}
Stringhini, G.; Kruegel, C.; and Vigna, G.
\newblock Detecting spammers on social networks.
\newblock In {\em ACSAC'10}.

\bibitem[\protect\citeauthoryear{Stutzman, Gross, and
  Acquisti}{2013}]{silentListeners}
Stutzman, F.; Gross, R.; and Acquisti, A.
\newblock 2013.
\newblock {Silent Listeners: The Evolution of Privacy and Disclosure on
  Facebook}.
\newblock {\em Journal of Privacy and Confidentiality} 4(2):2.

\bibitem[\protect\citeauthoryear{Tang \bgroup et al\mbox.\egroup
  }{2012}]{IndividualEmotions}
Tang, J.; Zhang, Y.; Sun, J.; Rao, J.; Yu, W.; Chen, Y.; and Fong, A. C.~M.
\newblock 2012.
\newblock {Quantitative Study of Individual Emotional States in Social
  Networks}.
\newblock {\em IEEE Transactions on Affective Computing} 3(2):132--144.

\bibitem[\protect\citeauthoryear{{Twitter, Inc.}}{2018}]{SearchAPI}
{Twitter, Inc.}
\newblock 2018.
\newblock {Twitter REST Public API documentation}.
\newblock
  \url{https://developer.twitter.com/en/docs/tweets/search/api-reference/get-search-tweets}.
\newblock Accessed: 2018-01-20.

\bibitem[\protect\citeauthoryear{Wang and Geng}{2009}]{wang2009behavioral}
Wang, L., and Geng, X.
\newblock 2009.
\newblock {\em Behavioral Biometrics for Human Identification: Intelligent
  Applications}.
\newblock IGI Global.

\bibitem[\protect\citeauthoryear{Wang \bgroup et al\mbox.\egroup
  }{}]{wang2012social}
Wang, G.; Mohanlal, M.; Wilson, C.; Wang, X.; Metzger, M.; Zheng, H.; and Zhao,
  B.~Y.
\newblock {Social Turing tests: Crowdsourcing sybil detection}.
\newblock In {\em {NDSS'13}}.

\end{thebibliography}
\end{small}
\end{quote}

\end{document}